\documentclass[11pt,amsfonts]{amsart}

\usepackage{fullpage}

\newtheorem{theorem}{Theorem}[section]

\newtheorem{definition}[theorem]{Definition}

\newtheorem{proposition}[theorem]{Proposition}
\newtheorem{remark}[theorem]{Remark}
\newtheorem{conjecture}[theorem]{Conjecture}

\def\endproof{\qed\medskip}
\def\blacksquare{\hbox to .60em {\vrule width .60em height .60em}}

\begin{document}

\title[]{Asymptotic Behavior of Future-Complete Cosmological Space-Times}

\author[]{Michael T. Anderson}

\thanks{Partially supported by NSF Grant DMS 0072591}

\maketitle

\begin{center}
Dedicated to Vince Moncrief on his $60^{\rm th}$ Birthday
\end{center}

\abstract
This work discusses the apriori possible asymptotic behavior to the future, for (vacuum) space-times which are geodesically complete to the future, and which admit a foliation by constant mean curvature compact Cauchy surfaces.
\endabstract

\setcounter{section}{0}

\section{Introduction.}
\setcounter{equation}{0}

   Let ({\bf M, g}) be a CMC cosmological space-time, i.e. a space-time with a compact constant mean curvature Cauchy surface $(\Sigma, g, K)$. The main focus will be on the vacuum case in $3+1$ dimensions although we will occasionally consider generalizations to non-negative energy conditions and higher dimensions.

 A fundamental issue in general relativity is to understand the global structure of ({\bf M, g}), and in particular the evolution of the geometry of the CMC foliation $\Sigma_{\tau}$ generated by the CMC slice $\Sigma .$ Singularities of ({\bf M, g}) will generally form in finite proper time, both to the future and to the past of $\Sigma$. Roughly, these may correspond either to big bang or big crunch singularities of the space-time as a whole, or to localized gravitational collapse within only parts of the space-time. The understanding of the mechanism and structure of such singularity formation is of course a central issue in general relativity.

 Here we concentrate instead on the simpler situation where there is no singularity formation (in finite proper time), say to the future of $\Sigma$. Thus, assume ({\bf M, g}) is geodesically complete (time-like and null) to the future of $\Sigma$. The basic issue then is to understand the asymptotic or long-time future behavior of the geometry of the CMC slices $\Sigma_{\tau}$ and of the full space-time ({\bf M, g}).

 Similar issues arise, and have been more fully investigated, in connection with globally hyperbolic, geodesically complete space-times with an asymptotically flat Cauchy surface, (purely radiative space-times). Thus, following Penrose, starting from an asymptotically flat initial data surface, one would like to understand the asymptotic structure of the resulting maximal globally hyperbolic space-time in terms of a conformal compactification leading to the structures of Scri; ${\mathcal I}$, ${\mathcal I}^{+}$ and ${\mathcal I}^{0}$. A great deal of work has been done and is ongoing in this direction, cf. [15], [21], [24] and references therein.

 The issues we consider then are cosmological analogues of this question. In principle, this should be a simpler situation to analyse, since there is no spatial infinity; compact slices are easier to deal with than non-compact slices.

 Vince Moncrief, together with Arthur Fischer, has begun a program to understand this issue, cf. [16]-[20] and further references therein. Independently, the author has also initiated such a program, cf. [1]. Both programs have a number of features in common and agree in certain special cases. In the end, they should also lead to the same overall description of the asymptotic behavior at future infinity. However, both the techniques and the point of view of these two approaches are rather different. The purpose of this paper in honor of Vince is to compare and comment on these two approaches.

\medskip

 The setting and assumptions we work with are the following. Let ({\bf M, g}) be a globally hyperbolic space-time, with a foliation ${\mathcal F}  = \{\Sigma_{\tau}\}$ of {\bf M} by compact, CMC Cauchy surfaces, parametrized by their mean curvature $\tau$. Unless stated otherwise, it is assumed that ({\bf M, g}) is a solution of the vacuum Einstein equations
\begin{equation} \label{e1.1}
{\bf Ric} = 0, 
\end{equation}
and that the foliation ${\mathcal F} $ is global in that ${\bf M}= {\bf M}_{{\mathcal F}}$, where ${\bf M}_{{\mathcal F}}$ is the union of the leaves $\Sigma_{\tau}$ in {\bf M}. It is also assumed that $\Sigma$ has no metric of positive scalar curvature. The Hamiltonian constraint then implies that $\tau  \leq 0$, and hence $\tau$ takes values within the interval $(-\infty, 0)$. Finally, assume that ({\bf M, g}) is geodesically complete to the future of an initial slice $\Sigma  = \Sigma_{\tau_{0}}$, so that runs over the full interval $\tau\in (\tau_{0}, 0)$. For surveys of prior work on CMC cosmological space-times, see [8], [25] or [28].

\section{Monotonicity Formulas.}
\setcounter{equation}{0}

 The spatial part of the evolution of the space-time may be described by the curve of metrics $g_{\tau}$ induced on the slices $\Sigma_{\tau}.$ Since $\Sigma_{\tau}$ is diffeomorphic to a fixed initial slice $\Sigma  = \Sigma_{\tau_{0}},$ this gives rise to a curve $g_{\tau}$ of metrics on a fixed 3-manifold $\Sigma$. Here we suppress the role of the shift diffeomorphisms $\Sigma \rightarrow  \Sigma_{\tau}$, since they will not play any significant role in the analysis.

 Apriori, the geometry of the curve of metrics $g_{\tau}$ induced on the 3-manifold $\Sigma $ could be quite arbitrary as $\tau  \rightarrow $ 0. In principle, there is not necessarily any limit. Even if such limits exist, they may apriori not have any special or simple features. Of course one would hope to prove that any limit has a much simpler structure than $(\Sigma_{\tau}, g_{\tau})$. In order to obtain some structure to the class of possible limits, it is very useful to find geometric quantities which behave monotonically w.r.t. the time evolution $\tau$, so that at least such quantities will have a well-defined limit.

\medskip

 In their study of the reduced phase space for the Einstein equations, i.e. a determination of the effective gravitational degrees of freedom, Fischer-Moncrief have found that the reduced system has a time-dependent scale-free Hamiltonian ${\mathcal H} ,$ which decreases along the Einstein flow, cf. [16], [18]. In fact, it turns out that ${\mathcal H} $ is just given by $|\tau|^{3}vol \Sigma_{\tau}.$ This leads to the monotonicity of the product:
\begin{equation} \label{e2.1}
{\mathcal H}  = |\tau|^{3}\cdot  vol \Sigma_{\tau} \downarrow , 
\end{equation}
as $\tau $ increases to 0. The behavior (2.1) follows from an analysis of the constraint equations and evolution equations for the slices $\Sigma_{\tau}$, (in a constant scalar curvature gauge), and so takes place on infinitesimal variations of the slices themselves. It does not bring into play the global structure of the space-time ({\bf M, g}), or how the slices sit within ({\bf M, g}).

 A different volume monotonicity for the slices $\Sigma_{\tau}$ was obtained in [1]. To define this, let $t_{\tau}$ be the Lorentzian distance from $\Sigma_{\tau}$ to $\Sigma_{\tau_{0}},$ i.e.
\begin{equation} \label{e2.2}
t_{\tau} = dist_{{\bf g}}(\Sigma_{\tau}, \Sigma_{\tau_{0}}) = \sup_{x\in\Sigma_{\tau}}dist_{{\bf g}}(x, \Sigma_{\tau_{0}}) > 0. 
\end{equation}
This is the maximal proper time between the slices $\Sigma_{\tau}, \Sigma_{\tau_{0}},$ and so of course depends on the global structure of the foliation ${\mathcal F} $ within ({\bf M, g}). Then [1, Cor.1.3] states that
\begin{equation} \label{e2.3}
\frac{vol \Sigma_{\tau}}{t_{\tau}^{3}} \downarrow , 
\end{equation}
i.e the ratio is monotone decreasing as $\tau$ increases. The proof of (2.3) rests only on elementary geometric principles; it follows by a simple combination of the Raychaudhuri equation and use of a geometric maximum principle dating back to Brill-Flaherty, cf. [11], [25] and [7].

 These monotonicity results contain on the one hand a number of similar features, and on the other hand, appear to be quite different. Actually, (2.3) follows by integration, over $\Sigma_{\tau}$ and in $\tau$, from its infinitesimal version
\begin{equation} \label{e2.4}
|\tau| \leq  \frac{3}{t_{\tau}}. 
\end{equation}
Suppose now $|\tau_{2}| \leq  |\tau_{1}|.$ Then substituting (2.4) in (2.3) gives
$$|\tau_{2}|^{3}vol \Sigma_{\tau_{2}} \leq  3\frac{vol \Sigma_{\tau_{1}}}{t_{\tau_{1}}^{3}} \sim  |\tau_{1}|^{3}vol \Sigma_{\tau_{1}}, $$
so that a very rough version of (2.1) follows from (2.4). However, there is no reason to believe that the proper time monotonicity (2.3) or (2.4) implies the monotonicity of the reduced Hamiltonian (2.1), and even less so vice-versa.

 Both monotonicity behaviors may be localized. Thus, if $U$ is a domain in {\bf M} obtained by flowing forward in time from some initial space-like domain $D \subset  \Sigma_{\tau_{1}}$, then (2.1) holds for the volume of the spatial slices $D_{\tau} \subset  \Sigma_{\tau}.$ The same holds with respect to (2.3); in fact (2.4) shows that (2.3) holds infinitesimally in space and time.

\medskip

 An important part of both monotonicity results is the implication on the stucture of the space-time when the product (2.1) or ratio (2.3) is constant. In this case, both approaches give identical results. Thus, if either (2.1) or (2.3) is constant on a $\tau$-interval $(\tau_{1}, \tau_{2})$, then the corresponding region of space-time ({\bf M, g}) bounded by $\Sigma_{\tau_{1}}$ and $\Sigma_{\tau_{2}}$ is isometric to a {\it  flat Lorentz cone}:
\begin{equation} \label{e2.5}
- dt^{2} + t^{2}g_{-1}, 
\end{equation}
where $g_{-1}$ is a hyperbolic metric on $\Sigma$ and $t = |\tau|^{-1}$. Note that this is the {\it simplest} vacuum space-time with a compact Cauchy surface, (except possibly the empty Minkowski quotient ${\mathbb R}\times T^{3}$). The space-time (2.5) is self-similar, with time evolution given by trivial rescalings of $g_{-1}$ and self-similarity generated by the vector field $\partial /\partial t$. 

\medskip

 All of the results discussed above hold in $n+1$ dimensional vacuum space-times, when the power $3$ in (2.1) and (2.3) is replaced by $n$; the coefficient $3$ in (2.4) also is replaced by $n$. For the monotonicity of the reduced Hamiltonian, this was noticed by Alan Rendall, and is discussed in [20]. For the proper time monotonicity, although not explicity noted, the proof is completely identical to that given in [1]. For the equality case in $n+1$ dimensions, in both cases the metric assumes the form of a Lorentzian cone on a Riemannian Einstein metric:
\begin{equation} \label{e2.6}
- dt^{2} + t^{2}g_{E}, 
\end{equation}
where $g_{E}$ is a Riemannian Einstein metric on $\Sigma$, i.e. $Ric_{g_{E}} = -(n-1)g_{E}$. 

 More generally, as noted in [1], the monotonicity (2.3) or (2.4) holds for space-times, (in $n+1$ dimensions), satisfying just the time-like convergence condition
\begin{equation} \label{e2.7}
Ric(T,T) \geq  0, 
\end{equation}
for time-like vectors $T$. In this case, the volume ratio (2.3) is constant on some $\tau$-interval if and only if the corresponding region of the space-time is of the form
\begin{equation} \label{e2.8}
-dt^{2} + t^{2}g^{-},
\end{equation}
where $g^{-}$ is a metric of negative Ricci curvature on $\Sigma$, $Ric_{g^{-}} \leq 0$. (We do not know if the monotonicity (2.1) holds in this more general setting).

\medskip

 It is curious that there exist two such distinct monotonic behaviors, both involving the volume of the CMC slices. Perhaps there is a family of such quantities, of which these two are special cases?

\section{Rescalings.}
\setcounter{equation}{0}

 Consider the geometry of the metrics $(\Sigma_{\tau}, g_{\tau})$ as $\tau  \rightarrow $ 0. Under the assumptions above that {\bf M} is geodesically complete to the future and ${\bf M} = {\bf M}_{{\mathcal F}}$, this forces $t_{\tau} \rightarrow  \infty$, (since there are no maximal hypersurfaces in ${\mathcal F}$). In all known situations, the metrics $g_{\tau}$ become locally flat and globally large, in that as $\tau \rightarrow 0$, one has the following estimates: 
\begin{equation} \label{e3.1}
|R_{\tau}| \rightarrow  0, \ vol B_{x_{\tau}}(1) \geq  v_{o}, \ vol_{g_{\tau}}\Sigma_{\tau} \rightarrow  \infty  \ {\rm and} \  diam_{g_{\tau}}\Sigma_{\tau} \rightarrow  \infty . 
\end{equation}
Here $|R_{\tau}|$ is the norm of the full curvature tensor of $g_{\tau}$, $x_{\tau}$ is any point in $\Sigma_{\tau}$, and $B_{x_{\tau}}(1)$ is the unit ball in $(\Sigma_{\tau}, g_{\tau})$ about $x_{\tau}$; $v_{o}$ is some fixed positive constant, depending on the initial data. It would be of interest to prove that the estimates (3.1) hold in general, (under the standing assumptions); it is hard to conceive of situations where any estimate in (3.1) fails.

 If (3.1) holds, then given any curve $x_{\tau}$ with $x_{\tau}\in\Sigma_{\tau},$ the Riemannian manifolds $(\Sigma_{\tau}, g_{\tau}, x_{\tau})$ based at $x_{\tau}$ converge to a flat metric on ${\mathbb R}^{3},$ (or possibly a flat quotient of ${\mathbb R}^{3}),$ (modulo diffeomorphisms). This is of course not very interesting; one has lost any understanding of the global geometry of $(\Sigma_{\tau}, g_{\tau})$ in the limit. The local geometry near any point becomes trivial, while the large-scale geometry escapes to infinity and is not detected in any limit. This situation is formally similar to taking a rescaling limit of any fixed metric $g$ at any point: as $\lambda \rightarrow 0$, the metrics $g_{\lambda} = \lambda^{-2}g$ converge, modulo diffeomorphisms, to the flat metric on ${\mathbb R}^{n}$, identified with the tangent space at the given point.

\medskip

 In order to obtain any meaningful global description of the future asymptotic behavior, one needs to rescale the slice metrics $g_{\tau}$ and the global space-time ({\bf M, g}). Exactly the same issue arises in describing the behavior at infinity of Minkowski space, or asymptotically flat space-times, at future space-like, null or time-like infinity.

 There are two very natural rescalings in view of the monotonicity formulas (2.1) and (2.3).

\smallskip
\noindent
(1) {\sf Mean curvature rescaling}:

 On each slice $\Sigma_{\tau},$ define the rescaled metric by
\begin{equation} \label{e3.2}
\widetilde g_{\tau} = \tau^{2}\cdot  g. 
\end{equation}
This corresponds to a constant rescaling of the space-time metric as 
\begin{equation} \label{e3.3}
{\bf \widetilde g}_{\tau} = \tau^{2}\cdot {\bf g}. 
\end{equation}
In the ${\bf \widetilde g}_{\tau}$ metric, the mean curvature of $\Sigma_{\tau}$ is now $- 1$, while the mean curvature of a general slice $\Sigma_{\mu}$ w.r.t. ${\bf \widetilde g}_{\tau}$ is $\widetilde \tau = \mu / |\tau|$. The monotonicity (2.1) just becomes the statement 
\begin{equation} \label{e3.4}
vol_{\widetilde g_{\tau}}\Sigma_{\tau} \downarrow , 
\end{equation}
i.e. the volume of $\Sigma_{\tau}$ is monotonically decreasing in the $\widetilde g_{\tau}$ metric. This rescaling is equivalent to the conformal volume rescaling used in [18], [19].

\smallskip
\noindent
(2) {\sf Proper Time Rescaling}:

 On each slice $\Sigma_{\tau}$, define the rescaled metric by
\begin{equation} \label{e3.5}
\bar g_{\tau} = t_{\tau}^{-2}\cdot  g. 
\end{equation}
This corresponds to a constant rescaling of the space-time metric as 
\begin{equation} \label{e3.6}
{\bf \bar g}_{\tau} = t_{\tau}^{-2}\cdot {\bf g}. 
\end{equation}
The scale-invariant bound (2.4) implies that the mean curvature of $\Sigma_{\tau}$ satisfies the bound $|\tau| \leq 3$. As above, the mean curvature of a slice $\Sigma_{\mu}$ w.r.t. ${\bf \bar g}_{\tau}$ is $\bar \tau = t_{\tau}\cdot \mu$. The volume monotonicity (2.3) becomes
\begin{equation} \label{e3.7}
vol_{\bar g_{\tau}}\Sigma_{\tau} \downarrow  . 
\end{equation}

 In both cases (3.3) and (3.6), one is rescaling the space-time ({\bf M, g}) by constants, so that the vacuum equations (1.1) or energy condition (2.7) is preserved. The asymptotic behavior thus involves considering limits of a family of metrics $\widetilde g_{\tau}$ or $\bar g_{\tau}$ on $\Sigma$ or ${\bf \widetilde g}_{\tau}$, ${\bf \bar g}_{\tau}$ on {\bf M}, as $\tau \rightarrow 0$. 

  In analogy to the Penrose compactification, one might consider instead a fixed global {\it  conformal}  rescaling of ({\bf M, g}); 
\begin{equation} \label{e3.8}
{\bf \widetilde g'} = \tau^{2}\cdot {\bf g}\ {\rm or} \  {\bf \bar g'} = t_{\tau}^{-2}\cdot {\bf g}. 
\end{equation}
These of course induce a constant spatial rescaling as in (3.2), (3.5), on each slice $\Sigma_{\tau}$. 

  Note however that in contrast to the asymptotically flat situation, the conformal metrics $({\bf M}, {\bf \widetilde g'})$ and $({\bf M}, {\bf \bar g'})$ are never conformal {\it  compactifications} of ({\bf M, g}). For instance, on the flat Lorentz cone (2.5), one has
\begin{equation} \label{e3.9}
{\bf \widetilde g'} = {\bf \bar g'} = - dlogt^{2} + g_{-1}, 
\end{equation}
which of course is not "compact"; one does not reach the boundary of the space-time in finite time in the conformal metric. The same phenomenon will hold on $({\bf M, g})$ itself. Because of this, there are no particular advantages to considering conformal rescalings as in (3.8); going to infinity in ({\bf M, g}) also requires going to infinity in $({\bf M, \widetilde g'})$ or $({\bf M, \bar g'})$.

\medskip

 While these two rescalings by mean curvature and proper time have similar formal properties, they may lead, at least apriori, to very different results. To explain this, recall that the Hawking-Penrose singularity theorem implies that the space-time ({\bf M, g}) comes to an end to the finite past of $\Sigma  = \Sigma_{\tau_{0}}$; thus, any past directed maximal time-like geodesic from $\Sigma$ has length at most $C/|\tau_{0}|$. The rescaling (3.6) essentially renormalizes the distance of $\Sigma_{\tau}$ to this initial singularity $S = \partial M$, in that
\begin{equation} \label{e3.10}
dist_{{\bf \bar g}_{\tau}}(\Sigma_{\tau}, S) \rightarrow  1, \ {\rm as} \ t_{\tau} \rightarrow  \infty . 
\end{equation}
Since the ${\bf \bar g}_{\tau}$-distance of $\Sigma_{\tau}$ remains bounded away from 0, the space-time and the geometry of slices $\bar g_{\tau}$ may have a limit as $\tau  \rightarrow 0$. (The existence of limits of $({\bf M, \bar g}_{\tau})$ and $(\Sigma , \bar g_{\tau})$ will be discussed more rigorously below; for now we just consider the situation formally). 

 On the other hand, this is not necessarily the case with respect to the rescaling (3.3). If
\begin{equation} \label{e3.11}
|\tau| <<  t_{\tau}^{-1}, \ {\rm as} \ \tau  \rightarrow  0, 
\end{equation}
then the rescaling (3.3) has the property that $dist_{{\bf \widetilde g}_{\tau}}(\Sigma_{\tau}, S) \rightarrow 0$, as $\tau \rightarrow 0$. In such a situation, there will be no limit space-time or limit geometry of $(\Sigma_{\tau}, \widetilde g_{\tau})$, since the geometry merges into that of the initial singularity $S$. When (3.11) holds, the rescaling (3.2)-(3.3) is much stronger than that of (3.5)-(3.6), and it pushes all the slices arbitrarily close to the singularity.

\medskip

 This argument suggests that the rescaling (3.6) is the correct one. Observe that rescalings that are uniformly bounded with respect to each another lead to the same limit behavior. Thus, given two rescaling functions $\lambda_{1}(\tau)$ and $\lambda_{2}(\tau)$, if $\lambda_{1}(\tau)/\lambda_{2}(\tau) + \lambda_{2}(\tau)/\lambda_{1}(\tau) \leq  \Lambda$, for some constant $\Lambda < \infty$, then the corresponding rescaled space-times will have the same geometric features; any limits will differ from each other by a uniformly bounded rescaling.

 Thus, if one can prove the opposite inequality to (3.11) necessarily holds, i.e. 
\begin{equation} \label{e3.12}
|\tau| \geq  \delta t_{\tau}^{-1}, 
\end{equation}
for some fixed $\delta > 0$, then the two rescalings will be equivalent.

 Finally, suppose for the sake of completeness, one considers rescalings $\lambda  = \lambda (\tau)$, $\hat g_{\tau} = \lambda^{2}g_{\tau}$, for which $\lambda(\tau) t_{\tau} \rightarrow \infty$. This will leave all slices at larger and larger distances to the initial singularity $S$. Any limit would then be a geodesically complete (vacuum) space-time, without singularities. The relation (2.4) implies that the limit has a foliation by maximal hypersurfaces. While one would expect any such limit to then be flat Minkowski space-time, and hence uninteresting, this is by no means clear. Apriori, it could be for instance a perturbation of Minkowski space, i.e. a Christodoulou-Klainerman space-time [15]. It would be very interesting if one could prove that Minkowski space-time, or a quotient of it, is the only possible limit in this case.

\section{Structure and Existence of Limits.}
\setcounter{equation}{0}

 For the remainder of the paper, we work with the proper-time rescaling (3.5)-(3.6) and consider the behavior of the family of rescaled space-times $({\bf M, \bar g_{\tau}})$ and the corresponding rescaled slices $(\Sigma_{\tau}, \bar g_{\tau})$ in the limit $\tau \rightarrow 0$. At a later point, we will then compare this with the mean curvature rescaling (3.2)-(3.3). To begin, we assume that limits exist, at least on a given sequence $\tau_{i} \rightarrow 0$. The deeper issues concerning the existence of such limits will be also discussed later.

\medskip

 The main tool to understand the limit structure is the volume monotonicity (2.1) or (2.3); since proper-time rescaling is being used, we work with (2.3). Suppose one has
\begin{equation} \label{e4.1}
vol_{\bar g_{\tau}}\Sigma_{\tau} \geq  v_{0} >  0, 
\end{equation}
as $\tau  \rightarrow 0$, for some arbitrary constant $v_{0} > 0$. Since the volume (4.1) is monotonically non-increasing, it follows that on any limit space-time $({\bf M_{0}, \bar g_{0}})$, the limit slices have constant volume ratio (2.3). The limit slices $\Sigma_{0}$ are parametrized by $\bar \tau$, where $\bar \tau = \lim_{\tau \rightarrow 0} t_{\tau}\cdot \tau$, so that the family $\Sigma_{0}(\bar \tau)$ is a family of CMC Cauchy surfaces in $({\bf M_{0}, \bar g_{0}})$ of constant mean curvature $\bar \tau$, cf. also (6.7)-(6.9) below. Thus, by the rigidity discussed in \S 2, the limit space-time $({\bf M_{0}, \bar g_{0}})$ is a flat Lorentz cone of the form (2.5), (in the vacuum case). The limit metric on the slices is a rescaling of the hyperbolic metric $g_{-1}$, and the space-time $({\bf M_{0}, \bar g_{0}})$ is self-similar. This conclusion is reached in both [17]-[19] and [1], although from different perspectives.

\medskip

 Thus, the structure of limits under the assumption (4.1) indeed appears to be very simple. The time evolution is asymptotic to a trivial, self-similar evolution. However, the situation is not quite so simple. The brief analysis above assumes that the limit slice $\Sigma_{0}$ is topologically the same as the initial slice $\Sigma$, so that the rescaled curve of metrics $\bar g_{\tau}$ converges to the limit hyperbolic metric $g_{-1}$ on $\Sigma$. (This of course then assumes that $\Sigma$ admits a hyperbolic metric). While this can occur, it does not necessarily occur. Even if $\Sigma$ admits a hyperbolic metric, the limit could be, for instance, a complete hyperbolic metric on the complement of a link in $\Sigma$ for which the volume of the complement goes to 0 as $\tau \rightarrow 0$ in the rescaled metric. Whether this can happen or not depends on whether the diameter of the rescaled slices, $diam_{\bar g_{\tau}}\Sigma_{\tau}$, remains bounded or becomes unbounded; see \S 5-\S 6 for further analysis. 

 If (4.1) does not hold, then one has
\begin{equation} \label{e4.2}
vol_{\bar g_{\tau}}\Sigma_{\tau} \rightarrow  0, \ {\rm as} \ \tau  \rightarrow  0, 
\end{equation}
so that the rescaled volume collapses. In this situation, there is of course no limit geometry to the slices. The curve of metrics $\bar g_{\tau}$ diverges in the space of Riemannian metrics, modulo diffeomorphisms. Consequently, there is no limit space-time. Nevertheless, we will see below in \S 5-\S 6 that in certain situations, this collapse can be resolved by passing to covering spaces of $\Sigma$ to obtain the existence of a rescaled limit space-time and CMC foliation.

\medskip

 The discussion above, although rigorous, is informal in that the existence of a rescaling limit has just been assumed. A deeper understanding must address the problem of determining situations or criteria when such limits actually exist.

 Ideally, one would like to develop criteria for the global existence for the Cauchy problem to the future of $\Sigma$, and the existence of rescaling limits, in terms of Cauchy data $(\Sigma, g, K)$ on the initial CMC Cauchy surface $\Sigma$. In general, this problem is extremely difficult, and well beyond current capabilities. However, in several more special situations, these issues have now recently been resolved to a large degree.

 Most notably, Vince Moncrief, in collaboration with L. Andersson [9] and Y. Choquet-Bruhat [14] respectively, has resolved these issues in the following two situations:

 (i). Let $\Sigma$ be a rigid hyperbolic 3-manifold, i.e. $\Sigma$ admits a hyperbolic metric $g_{-1}$ (of curvature -1), with no trace-free Codazzi tensors (i.e. infinitesimal conformally flat deformations). Then there is an $\varepsilon > 0$ such that if the Cauchy data $(\Sigma, g, K)$ satisfy 
$$||g - g_{-1}||_{H^{3}} \leq \varepsilon \ {\rm and} \ ||K + g_{-1}||_{H^{2}} \leq  \varepsilon,$$
then the vacuum space-time {\bf (M, g)} determined by $(\Sigma, g, K)$ is geodesically complete to the future of $\Sigma$, satisfies {\bf M} $= {\bf M}_{{\mathcal F}}$ and has a unique rescaling limit given by the Lorentz cone (2.5) on $(\Sigma , g_{-1})$.

 (ii). Let $\Sigma  = S^{1}\times \Sigma_{g}$, where $\Sigma_{g}$ is a surface of genus $g \geq  2$, and let $(\Sigma , g_{0}, K_{0})$ be homogeneous Cauchy data on $\Sigma$; thus $\Sigma$ is a $t = t_{0}$ slice in a Bianchi III space-time. Then there is an $\varepsilon > 0$ such that if the Cauchy data $(\Sigma, g, K)$ are polarized $S^{1}$ invariant, and satisfy 
$$||g - g_{0}||_{H^{2}(\Sigma_{g})} \leq  \varepsilon \ {\rm and} \ ||K - K_{0}||_{H^{1}(\Sigma_{g})} \leq  \varepsilon,$$
then the vacuum space-time {\bf (M, g)} determined by $(\Sigma, g, K)$ is geodesically complete to the future of $\Sigma$, and satisfies {\bf M} $= {\bf M}_{{\mathcal F}}$, (at least if the initial data satisfies a certain condition on the lowest eigenvalue of $\Delta$). The work in [14] does not specifically address rescaling limits, but one expects that there is collapse along the $S^{1}$ direction, which can be unwrapped in covering spaces to obtain a limit, cf. \S 5-\S 6; see also [29] for discussion of the future asymptotics of Bianchi III itself.

 As a further example, Gowdy space-times, at least in the polarized case, are known to be geodesically complete to the future, and satisfy {\bf M} $= {\bf M}_{{\mathcal F}}$, cf. [23] and references therein. It should also follow from the recent work of Ringstrom [30], cf. also [10], that such space-times have rescaling limits, after unwrapping collapse.

\medskip

 Returning to the general situation and to the standing assumption where ({\bf M, g}) is geodesically complete to the future of $\Sigma_{\tau_{0}}$ and {\bf M} $= {\bf M}_{{\mathcal F}}$, under what general conditions might a rescaling limit exist? It is reasonable to discuss this problem in terms of the space-time curvature {\bf R}, since this measures the strength of the gravitational field. 

 Thus, let $T$ be the future unit normal vector to the CMC slices $\Sigma_{\tau}$, and, as customary, measure the norm of the curvature w.r.t. $T$, i.e. 
\begin{equation} \label{e4.3}
|{\bf R}|^{2} = \sum{\bf R}_{ijkl}^{2}, 
\end{equation}
where the components run over an orthonormal basis $e_{i}$, $0 \leq  i \leq 3$, with $T = e_{0}$. (This is the electric-magnetic norm of the curvature w.r.t. the foliation ${\mathcal F}$). The norm measures in a certain sense the total gravitational force experienced by observers following the flow lines of $T$.

 If the space-time ({\bf M, g}) is to be geodesically complete to the future of $\Sigma_{\tau_{0}},$ one certainly expects $|{\bf R}|$ to remain bounded to the future of $\Sigma_{\tau_{0}}$, i.e. there should exist some constant $\Lambda  <  \infty$, (depending only on the Cauchy data on $\Sigma_{\tau_{0}})$, such that 
\begin{equation} \label{e4.4}
|{\bf R}| \leq  \Lambda . 
\end{equation}
This is an open problem. 

 When the metric is rescaled by $t_{\tau}^{-2}$ as in (3.6), the norm of the curvature of $({\bf M, \bar g_{\tau}})$ becomes
\begin{equation} \label{e4.5}
|{\bf \bar R}| = t_{\tau}^{2}|{\bf R}| . 
\end{equation}
Although apriori it may be possible to have a limit space-time where $|{\bf \bar R}|$ becomes unbounded, in practice there are no known situations where this occurs. Thus, we raise the following

\begin{conjecture}\label{c4.1}
 Let ({\bf M, g}) be a vacuum CMC cosmological space-time, geodesically complete to the future of $\Sigma_{\tau_{0}}$, with {\bf M} $= {\bf M}_{{\mathcal F}}$. Then there is a constant $\Lambda < \infty$, depending only on the initial data, such that
\begin{equation} \label{e4.6}
|{\bf \bar R}| = t_{\tau}^{2}|{\bf R}| \leq  \Lambda . 
\end{equation}
\end{conjecture}
Note that the quantity $t_{\tau}^{2}|{\bf R}|$ is scale-invariant. 

\begin{remark} \label{r 4.2.}
  {\rm It is worth pointing out that a certain converse to Conjecture 4.1 does hold.  Thus, if just (4.4) holds, and ${\bf M} = {\bf M}_{\mathcal F}$, then {\bf (M, g)} is geodesically complete to the future of $\Sigma$, cf. [1,Thm.0.1]. This gives a useful criterion for proving geodesic completeness. }
\end{remark}

 The bound (4.6) may still not enough to prove the existence of a rescaling limit with reasonable properties. The problem is related to how the slices $\Sigma_{\tau}$ sit inside the space-time ({\bf M, g}) when measured w.r.t. proper time. Thus, as noted in (3.10), in the rescaling (3.6), one has
\begin{equation} \label{e4.7}
\bar t_{\tau}(\Sigma_{\tau}, S) = \max_{x\in\Sigma_{\tau}} dist_{{\bf \bar g}_{\tau}}(x, S) \rightarrow  1, \ {\rm as} \ \tau  \rightarrow  0. 
\end{equation}
However, (4.7) does not preclude the possibility that certain parts of $\Sigma_{\tau}$ may wander down and approach the initial singularity $S$, i.e. one may have
\begin{equation} \label{e4.8}
\min_{x\in\Sigma_{\tau}} dist_{{\bf \bar g}_{\tau}}(x, S) \rightarrow  0, \ {\rm as} \ \tau  \rightarrow  0. 
\end{equation}
In this case, parts of the CMC slices on the limit all hit the singularity. (We know of no situations where this happens). In effect, one needs a bound of the form
\begin{equation} \label{e4.9}
\min_{x\in\Sigma_{\tau}} dist_{{\bf \bar g}_{\tau}}(x, S) \geq  \lambda  > 0, 
\end{equation}
so that $\Sigma_{\tau}$ lies in uniform proper-time annuli $(\lambda, 1)$ w.r.t. ${\bf \bar g}_{\tau}$.

 It has been proved in [1] that (4.9) holds, provided the bound (4.6) is strengthened somewhat to 
\begin{equation} \label{e4.10}
|{\bf \bar R}| + |\nabla{\bf \bar R}|= t_{\tau}^{2}|{\bf R}| + t_{\tau}^{3}|\nabla{\bf R}|\leq  \Lambda . 
\end{equation}
The reason the bound on the derivative of {\bf R} is needed is that part of the proof uses the Cauchy stability theorem. The weakest hypotheses to date on Cauchy stability require a bound on $\nabla{\bf R}$; a bound on {\bf R} is not currently known to suffice.

 We conjecture that also a stronger version of Conjecture 4.1 holds; namely under the same assumptions, (4.10) holds.

\begin{remark} \label{r 4.3.}
  {\rm We point out that under the bound (4.10), one has the following estimate, cf. [1,Thm.3.4]: there is a $\delta  = \delta (\Lambda ) > 0$ such that, for all $\tau$,
\begin{equation} \label{e4.11}
|\tau| \geq  \delta t_{\tau}^{-1}. 
\end{equation}
The bound (4.11) means that the mean curvature rescaling (3.2)-(3.3) and the proper-time rescaling (3.5)-(3.6) are equivalent, cf. (3.12). We know of no other (general) conditions implying (4.11). }
\end{remark}

\begin{remark} \label{r 4.4.}
  {\rm The bounds (4.6) and (4.10) involve the $L^{\infty}$ norm of the curvature tensor {\bf R}. The $L^{\infty}$ norm is very hard to control under the evolution of the space-time. In fact, regarding the results above and those to follow, (e.g. Theorem 6.2), the bounds (4.6) and (4.10) can be relaxed to $L^{2}$ bounds in place of $L^{\infty}$ bounds provided one has uniform control on the local volume, i.e. a lower bound on the volumes of all balls in $(\Sigma_{\tau}, \bar g_{\tau})$ of unit radius. }
\end{remark}

 Before proceeding further with the general discussion, let us note that the bound (4.10) does hold for the simplest models, namely the Bianchi space-times, provided the spatial velocity, measured by $K$, 
does not dominate too strongly the spatial geometry. Recall that the metric of a Bianchi space-time 
is given by 
$$g = - dt^{2} + \sum h_{ij}(t)\theta_{i}\cdot \theta_{j}. $$
\begin{proposition}\label{p 4.5.}
  Let ({\bf M, g}) be a Bianchi space-time, geodesically complete to the future, (so of any type except Bianchi $IX$). Suppose there exists $\lambda < \infty$ such that
\begin{equation}\label{e4.12}
|K|^{2} \leq \lambda |Ric|,
\end{equation}
where $K$ is the $2^{\rm nd}$ fundamental form and $Ric$ is the intrinsic Ricci curvature of the 
spatial slices $\{t = const\}$. Then the bound (4.10) holds on ({\bf M, g}), with $\Lambda = 
\Lambda(\lambda)$.
\end{proposition}
{\bf Proof:}
 We give an indirect proof, by contradiction, of the bound (4.10). This has the advantage of being conceptually simple, and avoids the involved computations of any specific model. The proof treats all cases simultaneously. In addition, this method of proof presents a very useful general point of view, applicable in many other situations. On the other hand, being by contradiction, the proof does not lead to any explicit bound; however, an explicit bound is irrelevant for our purposes. 

 Logically, the proof uses the results below describing the general situation regarding passing to limits of space-times. We will thus sketch all the ideas of the proof, and leave it to the reader to fill in the details for a rigorous argument, using the general theory described below in \S 5.

\medskip

 First, space-like slices of Bianchi space-times are homogeneous, so that {\bf R} and $\nabla{\bf R}$ are the same at all points on the slices $\{t = const\}$. Suppose, on some sequence 
$\tau_{i} \rightarrow 0$,
$$t_{{\tau}_{i}}^{2}|{\bf R}|_{\Sigma_{\tau_{i}}} \rightarrow  \infty, \ {\rm as} \ 
\tau_{i} \rightarrow  0. $$
Consider then the rescaled space-times ${\bf \hat g}_{i} = (|{\bf R}|_{\Sigma_{\tau_{i}}})^{-1}{\bf g}$, and the corresponding spatial metrics $\hat g_{i} = (|{\bf R}|_{\Sigma_{\tau_{i}}})^{-1}g_{\tau_{i}}$ on $\Sigma_{\tau_{i}}$. This rescaling has the effect that
\begin{equation} \label{e4.13}
|{\bf \hat R}| = 1, \ {\rm on} \ \Sigma_{\tau_{i}}. 
\end{equation}
The bounds (4.12) and (4.13) imply that the intrinsic curvature $\hat R$ of the slice $(\Sigma_{\tau_{i}}, \hat g_{i})$ is uniformly bounded, away from $0$ and $\infty$, for all $i$, (cf. [1,Prop.2.2]). Further, since the spatial geometry is homogeneous, the covariant derivatives $\nabla^{k}\hat R$ of the curvature of $\hat g_{i}$ are then also uniformly bounded, independent of $i$. (This is a general feature of homogeneous metrics, since the curvature and its derivatives are determined algebraically by the metric and the Lie bracket). 

 One may then take a rescaled limit $({\bf \hat M}_{0}, {\bf \hat g_{0}})$ with limit slice $(\hat \Sigma_{0}, \hat g_{0})$ on which 
\begin{equation} \label{e4.14}
|{\bf \hat R}| = 1 \ {\rm on} \ \hat \Sigma_{0}. 
\end{equation}
Here, we are assuming that one has convergence to a limit, and so no collapse. If the sequence $(\Sigma_{{\tau}_{i}}, \hat g_{i})$ collapses, the collapse may be unwrapped to obtain a convergent limit satisfying (4.14), cf. the end of \S 5 and Remark 6.4.

  The product (4.12) is scale-invariant, and hence it follows that the ${\bf \hat g}$-distance to the initial singularity diverges to $\infty$, as discussed at the end of \S 3. Thus, the limit is a Bianchi space-time which is geodesically complete, to the future {\it  and} to the past. However, it is easy to see that the only such Bianchi space-time is flat Minkowski space, or a quotient of it. This contradicts (4.14).

\medskip

 This proof shows that the bound (4.6) holds for all Bianchi space-times satisfying (4.12). The proof of (4.10), and analogues of (4.10) for all higher covariant derivatives of the curvature, follows in the same way; alternately, it follows from the remarks following (4.13).
{\endproof}

 Of course it would be interesting to know if this result also holds for small perturbations of any Bianchi initial data. It does hold for perturbations of hyperbolic initial data, by the work of Andersson-Moncrief [9]. It should also not be difficult to check if it holds for perturbations of Bianchi III as well as general Gowdy space-times, using [14] and [29]-[30]. 

\begin{remark}\label{r4.6}
{\rm In fact, the bound (4.10) holds for {\it all} Bianchi space-times satisfying (4.12), as one lets $t_{\tau} \rightarrow \infty$ or $t_{\tau} \rightarrow 0$. Thus, instead of diverging to future infinity, (4.10) holds as one approaches a big bang or big crunch singularity, where $t_{\tau} \rightarrow 0$. The proof is exactly the same as above, since all of the relevant quantities are scale-invariant. }
\end{remark}

\section{Spaces of Metrics with Bounded Curvature.}
\setcounter{equation}{0}

 In this section, we analyse the asymptotic behavior of general (vacuum) space-times ({\bf M, g}) satisfying the bound (4.6). We begin with an analysis of the behavior of the geometry of the space-like slices $(\Sigma_{\tau}, \bar g_{\tau})$ as $\tau  \rightarrow 0$.

 First, the bound (4.6) implies a bound on the intrinsic curvature $\bar R$ of $(\Sigma_{\tau}, \bar g_{\tau}):$
\begin{equation} \label{e5.1}
|\bar R| \leq  \Lambda'  = \Lambda' (\Lambda ), 
\end{equation}
see [1,Prop.2.2] for the proof. Further, the volume monotonicity (3.5) implies
\begin{equation} \label{e5.2}
vol_{\bar g_{\tau}}\Sigma_{\bar \tau} \leq  V, 
\end{equation}
where $V$ depends only on the initial data on $\Sigma_{\tau_{0}}$.

 Thus, one needs to understand the behavior of sequences or curves of metrics $(\Sigma_{\tau}, \bar g_{\tau})$ on a fixed 3-manifold $\Sigma$, as $\tau  \rightarrow 0$ or $\tau  = \tau_{i} \rightarrow 0$, under the bounds (5.1)-(5.2). What is the limiting behavior of such sequences or curves in the space of metrics on $\Sigma$?

 This is described by the Cheeger-Gromov theory, cf. [12], [13], [22], [2]-[4], and other references therein. Recall that on 2-surfaces, a sequence of constant curvature metrics may either converge, collapse, or form cusps, for example on $S^{2}, T^{2}$ or $\Sigma_{g}$ respectively, where $\Sigma_{g}$ is a surface of genus $g \geq 2$. The same basic trichotomy holds in dimension $3$, (and higher dimensions), under the bounds (5.1)-(5.2). For simplicity, we restrict the discussion to dimension $3$.

{\bf Convergence.}
 The space of Riemannian metrics on a (compact) 3-manifold $\Sigma$ such that
\begin{equation} \label{e5.3}
|R| \leq  \Lambda , \ vol \geq  v, \ diam \leq  D, 
\end{equation}
is precompact in the $C^{1,\alpha}$ and $L^{2,p}$ topologies. Thus, any sequence $\{g_{i}\}$ satisfying the bounds (5.1)-(5.2) has a subsequence, converging in the $C^{1,\alpha'}$, $\alpha' < \alpha$, and weak $L^{2,p}$ topology, to a limit $C^{1,\alpha}\cap L^{2,p}$ metric $g_{\infty}$ on $\Sigma$. Here and below, convergence is always understood to be modulo diffeomorphisms, (i.e. there is a sequence of diffeomorphisms $\phi_{i}$ such that $\phi_{i}^{*}g_{i}$ converges to a limit metric). The conditions (5.3) are of course invariant under diffeomorphisms.

\medskip

 When one of the global bounds in (5.3) fails, i.e. when $diam \rightarrow  \infty$, or $vol \rightarrow 0$, there is of course no limit metric on $\Sigma$ per se. There are then two further possibilities.

\medskip

{\bf Collapse.}
 Let $\{g_{i}\}$ be a sequence of Riemannian metrics on a (compact) 3-manifold $\Sigma$ satisfying
\begin{equation} \label{e5.4}
|R| \leq  \Lambda, \ vol_{g_{i}}\Sigma  \rightarrow  0, 
\end{equation}
Then $\Sigma$ is a graph manifold, and the metrics $g_{i}$ collapse $\Sigma$ to a lower dimensional space. In particular, there is no limit metric on $\Sigma$.

\medskip

 A graph manifold is a union of $S^{1}$ fibrations over surfaces, i.e. Seifert fibered spaces, glued together along toral boundary components. More precisely, a graph manifold $G$ has a decomposition into a disjoint union of Seifert fibered spaces $S = \{S_{i}\}$, with $\partial S_{i}$ a union of tori $\{T_{j}\}$. The manifold $G$ is then assembled by glueing (some of) the boundary tori together by toral automorphisms, i.e. elements of $SL(2, {\mathbb Z})$. Thus, $G$ decomposes as
\begin{equation} \label{e5.5}
G = S \cup  L, 
\end{equation}
where $S$ is a collection of Seifert fibered spaces and each component of $L$ is of the form $T^{2}\times I$: the components of $L$ glue together distinct toral boundary components in $S$. (One associates a graph to such a structure by assigning a vertex to each Seifert fibered space $S_{i}\in S$, and an edge joining $S_{k}$ to $S_{l}$ if a component of $L$ joins a boundary component of $S_{k}$ to a boundary component of $S_{l})$. The topology of graph manifolds is completely completely understood and classified, c.f. [34].

 This result implies that the conditions (5.4) can hold only under the very strong topological condition that $\Sigma $ is a graph manifold. The vast majority of 3-manifolds, (in a natural sense), are not graph manifolds.

 The collapse in (5.4) takes place by shrinking the $S^{1}$-fibers of the Seifert fibered components, and the $T^{2}$-fibers of the $L$ components to points. For example, let $S$ be an $S^{1}$ bundle over a surface $V$. Let $g_{V}$ be any metric on $V$, and let $\theta$ be a connection 1-form for the bundle. Then the curve of metrics 
\begin{equation} \label{e5.6}
g_{\varepsilon} = \varepsilon^{2}\theta^{2} + g_{V} 
\end{equation}
on $S$ collapses with bounded curvature as $\varepsilon \rightarrow 0$. Similarly on $I\times T^{2}$, the metrics $g_{\varepsilon} = dr^{2} + \varepsilon^{2}(d\theta_{1}^{2} + d\theta_{2}^{2})$ collapse the tori with bounded curvature to points as $\varepsilon \rightarrow 0$. 

\medskip

 Finally, we discuss the third possibility, the formation of cusps. This case is the most general and corresponds to a mixture of the two previous cases convergence/collapse; however no essentially new phenomena occur. To start, given a complete Riemannian manifold $(\Sigma, g)$, choose $\varepsilon > 0$ small, and let
\begin{equation} \label{e5.7}
\Sigma^{\varepsilon} = \{x\in\Sigma : volB_{x}(1) \geq  \varepsilon\}, \ \Sigma_{\varepsilon} = \{x\in\Sigma : volB_{x}(1) \leq  \varepsilon\}. 
\end{equation}
$\Sigma^{\varepsilon}$ is called the $\varepsilon$-thick part of $(\Sigma, g)$, while $\Sigma_{\varepsilon}$ is the $\varepsilon$-thin part.

 Now suppose $g_{i}$ is a sequence of complete Riemannian metrics on the manifold $\Sigma$. If the bounds (5.3) hold, then $(\Sigma, g_{i})$ satisfies $\Sigma  = \Sigma^{\varepsilon}$, for a suitable $\varepsilon  = \varepsilon (k, v, D)$. (This follows easily from the standard Bishop-Gromov volume comparison theorem, cf. [27]). On the other hand, if for any given $\varepsilon > 0$ and $i = i(\varepsilon)$ sufficiently large, one has $(\Sigma_{\varepsilon})_{g_{i}} = \Sigma$, then the sequence $(\Sigma, g_{i})$ is collapsing, (athough the total volume may not be forced to go to 0).

 The only remaining possibility is that there exist points $x_{i}$ and $y_{i}$ in $\Sigma$ such that
\begin{equation} \label{e5.8}
vol B_{x_{i}}(1) \geq  \varepsilon_{o}, \ vol B_{y_{i}}(1) \rightarrow  0, 
\end{equation}
for some $\varepsilon_{o} > 0$. Observe that the volume comparison theorem implies that $dist_{g_{i}}(x_{i}, y_{i}) \rightarrow  \infty$ as $i \rightarrow  \infty$, so that these different behaviors become further and further distant as $i \rightarrow  \infty$.

 Define a {\it  cusp} to be a Riemannian 3-manifold $(N, g)$ where $N$ is open, and $g$ is a complete, finite volume, $C^{1,\alpha} \cap L^{2,p}$ metric on $N$ with curvature bounded in $L^{p}$. Outside a sufficiently large compact set, $N$ is a graph manifold (with toral boundary components).

\medskip

{\bf Cusps.}
 Let $\Sigma$ be a $3$-manifold and $g_{i}$ a sequence of metrics on $\Sigma$ satisfying (5.1)-(5.2) and (5.8). Choose also a sequence $\varepsilon_{i} \rightarrow 0$. Then, in a subsequence, each non-collapsing component of $\Sigma^{\varepsilon_{i}}$ converges to a cusp $(N_{j}, g_{\infty})$, uniformly on compact sets, while each component of $\Sigma_{\varepsilon_{i}}$ or collapsing component of $\Sigma^{\varepsilon_{i}}$ is a graph manifold $G_{k}$, collapsing to a lower dimensional space as $i \rightarrow  \infty$. The convergence is as above, i.e. in the $C^{1,\alpha}$ and weak $L^{2,p}$ topologies.

 The number of limit cusps and graph manifold components may be (countably) infinite, and
\begin{equation} \label{e5.9}
\sum_{j}vol_{g_{\infty}}N_{j} \leq  V = \overline{\lim} vol_{g_{i}}\Sigma. 
\end{equation}
Each cusp $N_{j}$ weakly embeds in $\Sigma$, in the sense that any compact domain of $N_{j}$ embeds as a domain in $\Sigma$. Formally, one may write
\begin{equation} \label{e5.10}
\Sigma  = N \cup  G, 
\end{equation}
where $N$ is the union of the cusps and $G$ is the union of the graph manifold components. Thus, if $G = \emptyset$, then one is in the (pure) convergence situation and hence $N = \Sigma$, while if $N = \emptyset$, then one is in the (pure) collapse situation, with $G = \Sigma$.

 In contrast to the (pure) collapse situation, there are no topological restrictions for a sequence $(\Sigma , g_{i})$ to form cusps. The decomposition (5.10) is not necessarily related to the topology of $\Sigma$. For instance, if $\Sigma$ is any closed 3-manifold, one may choose $G$ to be a tubular neighborhood of any link in $\Sigma$ and set $N = \Sigma \setminus G$. It is not difficult to construct metrics on $\Sigma$ satisfying (5.1)-(5.2) and (5.8) which converge to cusps on $N$ and collapse $G$. In this generality, $N$ could have an infinite number of components.

\medskip

 This completes the discussion of the general trichotomy. There is one further important feature of collapse in dimension $3$. Thus, suppose in the decomposition (5.5) of a graph manifold $G$, no component $S_{i}$ of $S$ is a spherical space form $S^{3}/\Gamma$, or is a solid torus $D^{2}\times S^{1}$. The former case occurs of course only if $G = S^{3}/\Gamma$. Then one knows from $3$-manifold topology that $\pi_{1}(S_{j})$ and $\pi_{1}(L_{k})$ inject in $\pi_{1}(G)$ for all components $S_{j}$ and $L_{k}$ of $S$ and $L$. Further, the $S^{1}$ fibers of each $S_{j}$ also inject in $\pi_{1}$. In particular, if $G = S$ is a closed Seifert fibered space, then $\pi_{1}(S^{1})$ always injects in $\pi_{1}(S)$ unless $S = S^{3}/\Gamma$. 

 This has the implication that one may pass to covering spaces to {\it unwrap the collapse}. As discussed above, the collapse takes place by shrinking the $S^{1}$ and $T^{2}$ fibers in the $S$ and $L$ components to points, respectively. One can unwrap such small $S^{1}$'s or $T^{2}$'s by passing to sufficiently large covering spaces, so they are no longer small. One sees this easily in the specific example of collapse in (5.6), but this structure holds in general. Thus, by passing to such local covering spaces, one can obtain convergence to limits; we refer to [1], [4] for further discussion and applications of this fact.

\section{Asymptotics and Geometrization of 3-Manifolds.}
\setcounter{equation}{0}

 This classification of the limiting behavior of sequences of metrics with bounded curvature on 3-manifolds is closely related with Thurston's picture for the topological classification of 3-manifolds, see [31], [32]. To describe this in somewhat more detail, we need the following definition.

\begin{definition}\label{d6.1}
 Let $\Sigma$ be a closed, oriented and connected 3-manifold. A {\sf  weak}  geometrization of $\Sigma$ is a decomposition
\begin{equation} \label{e6.1}
\Sigma  = H \cup  G, 
\end{equation}
where $H$ is a finite collection of complete connected {\sf hyperbolic manifolds} of finite volume embedded in $\Sigma$ and $G$ is a finite collection of connected {\sf graph manifolds} embedded in $\Sigma$. The union is along a finite collection of embedded tori ${\mathcal T}  = \cup T_{i}$, ${\mathcal T}  = \partial H = \partial G$.

 A {\sf  strong}  geometrization of $\Sigma$  is a weak geometrization as above, for which each torus $T_{i} \in  {\mathcal T}$ is incompressible in $\Sigma$, i.e the inclusion of $T_{i}$ into $\Sigma$ induces an injection of fundamental groups.
\end{definition}
 Of course, it is possible that the collection ${\mathcal T}$ of tori dividing $H$ and $G$ in (6.1) is empty, in which case weak and strong geometrizations are the same. In such a situation, $\Sigma$ is then either a closed hyperbolic manifold or a closed graph manifold. Note the similarity of (6.1) with (5.10). For a strong geometrization, the decomposition (6.1) is unique up to isotopy, c.f. [32], (or [5]), but this is far from being the case for a weak geometrization, for the same reasons as discussed following (5.10).

\medskip

 We may now apply the general structural results on sequences of metrics in \S 5 to any sequence $(\Sigma_{\tau_{i}}, \bar g_{\tau_{i}})$. This leads to the following result [1] describing in general the possible asymptotic behavior of ({\bf M, g}) at future infinity, at least under the bound (4.10).

\begin{theorem} \label{t 6.2.}
  Let ({\bf M, g}) be a CMC cosmological vacuum space-time, with {\bf M} $= {\bf M}_{{\mathcal F}}$, and satisfying (4.10), i.e.
\begin{equation} \label{e6.2}
|{\bf \bar R}| + |\nabla{\bf \bar R}|\leq  \Lambda . 
\end{equation}
Then:

 (I). (M, g) is geodesically complete to the future of the initial data surface $\Sigma_{\tau_{0}}$.

 (II). Given any sequence $\tau_{i} \rightarrow 0$, a subsequence of the slices $(\Sigma_{\tau_{i}}, \bar g_{\tau_{i}})$ converges to a weak geometrization of $\Sigma$. Thus, the metrics $\bar g_{\tau_{i}}|_{H}$ converge to the complete hyperbolic metric on $H$, while $\bar g_{\tau_{i}}|_{G}$ collapses the graph manifolds in $G$ to a lower dimensional space.
\end{theorem}

{\bf Idea of Proof:}  The main statement here is (II). The proof of (I) follows just from a bound on $|{\bf R}|$, cf. Remark 4.2. Thus, let 
\begin{equation} \label{e6.3}
V_{\infty} = \lim_{\tau\rightarrow 0} t_{\tau}^{-3}vol_{g_{\tau}}\Sigma_{\tau} = \lim_{\tau\rightarrow 0}vol_{\bar g_{\tau}}\Sigma_{\tau}, 
\end{equation}
be the limit of the rescaled volumes. 

  The simplest situation is when 
\begin{equation} \label{e6.4}
V_{\infty} > 0 \  {\rm and} \  diam_{\bar g_{\tau}}(\Sigma_{\tau}, \bar g_{\tau}) < D, 
\end{equation}
for some arbitrary constant $D <  \infty$, and $\tau  = \tau_{i}$. In this situation, the bounds (5.3) hold, and so one has convergence to a limit, (in a subsequence). The volume monotonicity (2.3) implies that any limit is hyperbolic, i.e. has a metric of constant negative curvature, (which can be scaled to -1). In these circumstances, the limit is unique, so that the full curve of metrics $\bar g_{\tau}$ converges to the hyperbolic metric. 

 If instead
\begin{equation} \label{e6.5}
V_{\infty} > 0 \ {\rm but} \  diam_{\bar g_{\tau}}(\Sigma_{\tau}, \bar g_{\tau}) \rightarrow  \infty , 
\end{equation}
then one is in the cusp situation (5.8), (or possibly the collapse situation). Using the local version of the monotonicity (2.3) discussed in \S 2, one finds that the limits based at any sequence of points $x_{i}$ as in (5.8) converge to a complete hyperbolic cusp $H_{k},$ as in (6.1). On the other hand, the geometry at base points $y_{i}$ as in (5.8) collapses, and their is no limit metric. This situation corresponds to the decomposition (6.1).

 Finally, if
\begin{equation} \label{e6.6}
V_{\infty} = 0, 
\end{equation}
then the metrics $(\Sigma_{\tau}, \bar g_{\tau})$ collapse; in particular, $\Sigma$ must be a graph manifold in this case.

\medskip

  Theorem 6.2 gives a close relationship between the possible future asymptotic behavior of the space-time ({\bf M, g}) and geometrization of 3-manifolds. For this relationship to be more meaningful, one would need to know that the asymptotic geometry of the slices $(\Sigma_{\tau}, \bar g_{\tau})$ induces a strong geometrization of $\Sigma$; see [1] for further remarks on this. 

\medskip

   Although Theorem 6.2 as stated describes only the future spatial asymptotic geometry, it extends easily to a statement regarding the asymptotics of the full space-time ({\bf M, g}). Thus, the CMC foliation ${\mathcal F}$ gives a $3+1$ decomposition of ({\bf M, g}) as
\begin{equation} \label{e6.7}
{\bf g} = -\alpha^{2}d\tau^{2} + g_{\tau} = -\alpha^{2}(\frac{d\tau}{dt_{\tau}})^{2}dt_{\tau}^{2} + g_{\tau}. 
\end{equation}
The quantity $\beta = \alpha(\frac{d\tau}{dt_{\tau}})$ is scale-invariant and using (2.4) and (4.11), one has uniform bounds
\begin{equation} \label{e6.8}
k \leq \alpha (\frac{d\tau}{dt_{\tau}}) \leq K,
\end{equation}
for some constants $k, K$ independent of $\tau$. The estimate (6.8) of course requires the bound (6.2).

  Thus, given a sequence $\tau_{i} \rightarrow 0$ as in Theorem 6.2, one has, for a subsequence, a limit space-time $({\bf M_{0}, \bar g_{0}})$ of the form 
\begin{equation} \label{e6.9}
{\bf g_{0}} = -\alpha_{0}^{2}d\bar \tau^{2} + g_{\bar \tau} = -\beta_{0}^{2}d\bar t_{\tau}^{2} + g_{\bar t_{\tau}}.
\end{equation}
Here, one easily computes that $\bar \tau = \lim \frac{\tau}{\tau_{i}}$ and $\alpha_{0} = \lim \tau_{i}^{2}\alpha$, while $\bar t_{\tau} = \lim t_{\tau}/t_{\tau_{i}}$ and $\beta_{0} = \lim \beta_{i}$, where $\beta_{i} = \alpha(\frac{d\tau}{dt_{\tau_{i}}})$.

  In the hyperbolic region $H$ of (6.1), the metric ${\bf g_{0}}$ is a Lorentz cone, as in (2.5). In the collapsed graph manifold region $G$, the metric ${\bf g_{0}}$ in (6.9) is defined only if the collapse can be unwrapped in covering spaces, as discussed at the end of \S 5.

\begin{remark} \label{r 6.3.}
  {\rm By Proposition 4.4, all the (Class A) Bianchi space-times excluding the recollapsing models, (Bianchi IX and Kantowski-Sachs), satisfy the hypotheses of Theorem 6.2, and hence all satisfy the conclusions. Since the spatial geometries are homogeneous, cusps cannot form. Thus, the decomposition (6.1) is necessarily pure in this case. 

  The case $\Sigma  = H$, i.e.  $G = \emptyset$, corresponds to Bianchi $V$, $VII_{h}$, while the case $\Sigma  = G$, i.e. $H = \emptyset$, corresponds to all other cases. Of course, the collapse can be unwrapped to obtain a limit geometry as in (6.9), for all expanding Bianchi models. } 
\end{remark}

\begin{remark} \label{r 6.4.}
{\rm While the main concern above has been with vacuum space-times, all of the results above hold for space-times satisfying the time-like convergence condition (2.7). Of course, one would then not obtain limits which are flat Lorentz cones, as in (2.5); instead, non-collapsed parts of the limit have the form (2.8). 

  In addition, since the Cheeger-Gromov theory holds in arbitrary dimensions, there are natural analogues of these results in any dimension. }
\end{remark}

\section{Relations with the Sigma Constant.}
\setcounter{equation}{0}

 Let ${\mathcal M}_{-1}$ denote the space of metrics on $\Sigma$, modulo diffeomorphisms, of constant scalar curvature $-1$; we recall that $\Sigma$ is assumed to admit no metric of positive scalar curvature. The Sigma constant $\sigma (\Sigma)$ is defined by
\begin{equation} \label{e7.1}
\sigma (\Sigma ) = -  \inf_{g\in{\mathcal M}_{-1}}(vol_{g}\Sigma )^{2/3}. 
\end{equation}
This is a topological invariant of the 3-manifold $\Sigma$. The definition of $\sigma (\Sigma)$ extends to 3-manifolds which admit positive scalar curvature metrics, but this will be irrelevant here. The definition (7.1) is of course equivalent to the scale-invariant definition 
\begin{equation} \label{e7.2}
\sigma (\Sigma ) = \inf_{g\in{\mathcal C}}R_{g}(vol_{g}\Sigma )^{2/3}, 
\end{equation}
where ${\mathcal C}$ is the space of constant scalar curvature metrics, and $R_{g}$ is the scalar curvature of $g$.

 Fischer and Moncrief have shown [16]-[19], that the infimum of the reduced Hamiltonian ${\mathcal H}$ for vacuum space-times over the reduced phase space ${\mathcal M}_{-1}$ is determined by the Sigma constant:
\begin{equation} \label{e7.3}
\inf_{{\mathcal M}_{-1}}{\mathcal H}  = (\tfrac{3}{2}|\sigma (\Sigma )|)^{3/2}. 
\end{equation}

 Here we give an alternate elementary proof of (7.3). The Hamiltonian constraint equation on $(\Sigma_{\tau}, g_{\tau})$ in the vacuum case gives
\begin{equation} \label{e7.4}
R = |K|^{2} - \tau^{2} = |K_{0}|^{2} -  \tfrac{2}{3}\tau^{2} \geq  -  \tfrac{2}{3}\tau^{2}, 
\end{equation}
where $K_{0}$ is the trace-free extrinsic curvature. 

 Let $\hat g_{\tau}$ be the Yamabe metric, i.e. constant scalar curvature metric, conformal to $g_{\tau}$ with the same volume. Thus, $\hat g_{\tau} = \phi^{4}g_{\tau},$ where $\phi$ is a positive function satisfying the Yamabe equation
\begin{equation} \label{e7.5}
-8\Delta \phi  + R\phi  = \phi^{5}\hat R, 
\end{equation}
on $(\Sigma_{\tau}, g_{\tau})$, where $\hat R$ is a negative constant. Evaluating (7.5) at a point realizing the minimum of $\phi$ then gives
\begin{equation} \label{e7.6}
\hat R \geq  -\frac{2}{3}\tau^{2}, 
\end{equation}
i.e. the Yamabe metric $\hat g$ has scalar curvature $\geq  -\frac{2}{3}\tau^{2}$. Setting $\mu  = \frac{2}{3}\tau^{2}$, if one rescales again to make $R = -\mu$, then the volume decreases further. Hence, one has
\begin{equation} \label{e7.7}
|\tau|^{3}vol_{g_{\tau}}\Sigma_{\tau} \geq  |\tau|^{3}\inf_{{\mathcal M}_{-\mu}}vol_{g}\Sigma  = |\tau|^{3}\mu^{-3/2}\inf_{{\mathcal M}_{-1}}vol_{g}\Sigma , 
\end{equation}
where the last equality again follows from a simple rescaling. Combining (7.7) with the definition (7.1) gives, for all $\tau$,
\begin{equation} \label{e7.8}
{\mathcal H}  = |\tau|^{3}vol_{g_{\tau}}\Sigma_{\tau} \geq  (\tfrac{3}{2}|\sigma (\Sigma)|)^{3/2}. 
\end{equation}
On the other hand, for a fixed $\tau$, say $\tau = -(\frac{3}{2})^{1/2}$, one can find Yamabe metrics $\hat g$ with $\hat R = -1$ of volume arbitrarily close to $|\sigma(\Sigma)|^{3/2}$. Setting $K_{0} = 0$, one has then a solution of the Hamiltonian and momentum constraints, and hence $(\Sigma, \hat g, K)$ generates some maximal space-time ({\bf M, g}). This gives the equality (7.3). 

  A similar result holds for the volume monotonicity (2.3). Thus, from (2.4) and (7.8), one has
\begin{equation} \label{e7.9}
\inf \frac{vol \Sigma_{\tau}}{t_{\tau}^{3}} \geq (\tfrac{1}{6}\sigma(\Sigma))^{3/2},
\end{equation}
where the inf is over all globally hyperbolic vacuum space-times having $\Sigma$ as a Cauchy surface. It is an open question whether equality holds in (7.9); this involves more global issues of how the initial surface $\Sigma$ sits inside the full space-time ({\bf M, g}).

\begin{remark} \label{r 7.1.}
  {\rm The argument above only requires the inequality (7.4). The general Hamiltonian constraint equation, (in non-vacuum), reads
\begin{equation} \label{e7.10}
R = |K|^{2} - \tau^{2} + 2{\bf Ric}(N,N) + {\bf R}, 
\end{equation}
where $N$ is the time-like unit normal to the CMC foliation; here {\bf Ric} and {\bf R} are the space-time Ricci and scalar curvature. Since the Einstein field equations give $2{\bf Ric}(N,N) + {\bf R} = 2T(N, N)$, where $T$ is the energy-momentum tensor, it follows that (7.3) holds for space-times satisfying just the weak energy condition. }
\end{remark}

\begin{remark} \label{r 7.2.}
  {\rm Exactly the same proof of (7.3) holds in all dimensions, with the standard modification of the Yamabe equation (7.5) in higher dimensions. However, the Sigma constant is no longer closely associated with the topology of higher dimensional manifolds. For instance, $\sigma (\Sigma) \geq 0$ for all simply-connected manifolds in dimensions $\geq 5$, cf. [26]. 

 For further discussion of the relation between $\sigma(\Sigma)$ and geometrization of the 3-manifold $\Sigma$, we refer to [3], [5] or [6]. }
\end{remark}

{\bf Ackowledgement} I would like to thank Hans Ringstr\"om for remarks and correspondence regarding 
the structure of Bianchi and Gowdy space-times related to the issues discussed here.

\medskip

{\it Note.} Since this paper was written in March, 2003, Grisha Perelman [35] has announced a proof 
of Thurston's Geometrization Conjecture [32]

\bibliographystyle{plain}

\bigskip
\begin{center}
March, 2003/December, 2003
\end{center}

\medskip
\noindent
\address{Department of Mathematics\\
S.U.N.Y. at Stony Brook\\
Stony Brook, New York 11794-3651}\\
\email{anderson@math.sunysb.edu}

\end{document}